\documentclass[conference]{IEEEtran}
\IEEEoverridecommandlockouts
\usepackage{cite}
\usepackage{amsmath,amssymb,amsfonts}
\usepackage{algorithmic}
\usepackage{graphicx}
\usepackage{textcomp}
\usepackage{xcolor}
\def\BibTeX{{\rm B\kern-.05em{\sc i\kern-.025em b}\kern-.08em
    T\kern-.1667em\lower.7ex\hbox{E}\kern-.125emX}}

\usepackage{comment}
\usepackage{hyperref}
\usepackage{booktabs}
\usepackage{adjustbox}
\usepackage{placeins}

\begin{document}

\title{ConGaIT: A Clinician-Centered Dashboard for Contestable AI in Parkinson’s Disease Care}

\author{\IEEEauthorblockN{Phuc Truong Loc Nguyen}
\IEEEauthorblockA{
\textit{Friedrich-Alexander University of Erlangen-Nuremberg}\\
Erlangen, Germany \\
loc.pt.nguyen@fau.de\\
0009-0003-4254-0750}
\and
\IEEEauthorblockN{Thanh Hung Do}
\IEEEauthorblockA{
\textit{Friedrich-Alexander University of Erlangen-Nuremberg}\\
Erlangen, Germany \\
hung.t.do@fau.de\\
0009-0005-2025-5080}
}

\maketitle

\begin{abstract}
AI-assisted gait analysis holds promise for improving Parkinson’s Disease (PD) care, but current clinical dashboards lack transparency and offer no meaningful way for clinicians to interrogate or contest AI decisions. We present ConGaIT (Contestable Gait Interpretation \& Tracking), a clinician-centered system that advances Contestable AI through a tightly integrated interface designed for interpretability, oversight, and procedural recourse. Grounded in HCI principles, ConGaIT enables structured disagreement via a novel \textit{Contest \& Justify} interaction pattern, supported by visual explanations, role-based feedback, and traceable justification logs. Evaluated using the Contestability Assessment Score (CAS), the framework achieves a score of 0.970, demonstrating that contestability can be operationalized through human-centered design in compliance with emerging regulatory standards. A demonstration of the framework is available at \href{https://github.com/hungdothanh/Con-GaIT}{https://github.com/hungdothanh/Con-GaIT}.
\end{abstract}


\section{Introduction}
Parkinson’s Disease (PD) is a progressive neurodegenerative condition marked by motor impairments such as bradykinesia, gait freezing, and instability \cite{jankovic2008parkinson}. While symptom progression demands continuous monitoring and timely intervention, clinical assessments remain infrequent and subjective. Recent advances in foot-mounted inertial sensors enable continuous gait tracking in natural settings, and AI models can now extract valuable insights for disease staging \cite{Zhang2024}. Yet, these technical advances often fail in practice due to the lack of interfaces that support interpretation, interaction, and clinical integration \cite{He2019}.

Existing PD dashboards offer limited usability, often reducing complex AI outputs to static or opaque displays with little contextual support or clinician feedback. This gap undermines trust, restricts adaptability to atypical cases, and violates the transparency and oversight standards of the EU AI Act \cite{eu2024ai} and GDPR Article 22 \cite{eu_gdpr_2016_misc}. From an HCI standpoint, there is a growing demand for systems that promote interpretability, traceability, and collaborative decision-making.

We introduce a clinician-centered PD dashboard that operationalizes Contestable AI (CAI) principles across three core modules: \textbf{Gait Session Summary}, \textbf{Treatment Trend View}, and \textbf{Predictive Insight and Explanation}. Each module is enhanced for contextual visualization, intuitive interaction, and structured feedback. Clinicians can challenge outputs using defined argument types via a role-based interface, supported by a vision language model that provides normative justifications grounded in clinical criteria such as Hoehn and Yahr \cite{hoehn1967parkinsonism}. The framework is evaluated using the Contestability Assessment Score (CAS) \cite{moreira2025explainable}, which quantifies explainability, traceability, and user empowerment. This work aims to advance HCI-centered, regulation-aligned clinical AI design.

\section{Background and Related Work}
Parkinson’s Disease (PD) care is shifting from episodic clinical visits to continuous, sensor-driven monitoring \cite{s23104957}. QDG-Care \cite{hoffman2024comprehensive} exemplifies this trend by using Quantitative Digitography to remotely assess motor symptoms. Built on HCI principles, it integrates with EHRs via SMART on FHIR \cite{mandel2016smart} and presents a multi-layered dashboard combining high-level metrics with color-coded drill-downs aligned to normative baselines. Clinicians can align symptom metrics with medication timing to evaluate treatment response. The system uses statistical models and XGBoost for score generation. However, it lacks transparency into decision logic and offers no way for users to probe or challenge AI-driven outputs.

Explainable AI (XAI) helps address these gaps by clarifying how models make predictions. XAI techniques vary in scope, timing, and output, ranging from feature attributions to visual or textual explanations \cite{molnar2020interpretable,NGUYEN2025102782,ijcai2024p1025,nguyen2025heart2mind}. In clinical settings, they enhance trust by revealing feature contributions or uncertainty estimates \cite{weber2023beyond}. Yet most implementations remain static and non-interactive, requiring manual interpretation and offering no mechanism for feedback or correction. This limits their value in collaborative, HCI-driven domains like healthcare, where explanations must be cognitively aligned, actionable, and interface-integrated \cite{herrera2025opacity,speckmann2025ixaii}.

Contestable AI (CAI) extends XAI by embedding procedural structures that allow users to challenge, revise, and justify AI outcomes \cite{nguyen2025human}. Defined as a system-level property rooted in procedural justice \cite{alfrink2023contestable}, CAI emphasizes user control, traceability, and dialogic interaction. A core distinction is between explanation (how a result was generated) and justification (why it is normatively valid) \cite{dignum2025contesting}. Contestation is operationalized through deliberation argument trees, framing structured exchanges between the clinician (beneficiary) and the AI system or delegate (representative). Argument types, such as supposition, association, and perception, guide this dialogue. Third parties like ethics boards may intervene when disputes escalate. The CAS \cite{moreira2025explainable} evaluates how well systems support such engagement across eight criteria spanning four pillars: human-centered, technical, legal, and organizational.

To the best of our knowledge, no prior work has operationalized CAI within clinician-centered dashboards for PD care to unify AI support, explainability, and human oversight.

\section{Proposed Framework}

\begin{figure}[t!]
    \centering
    \includegraphics[width=1\linewidth]{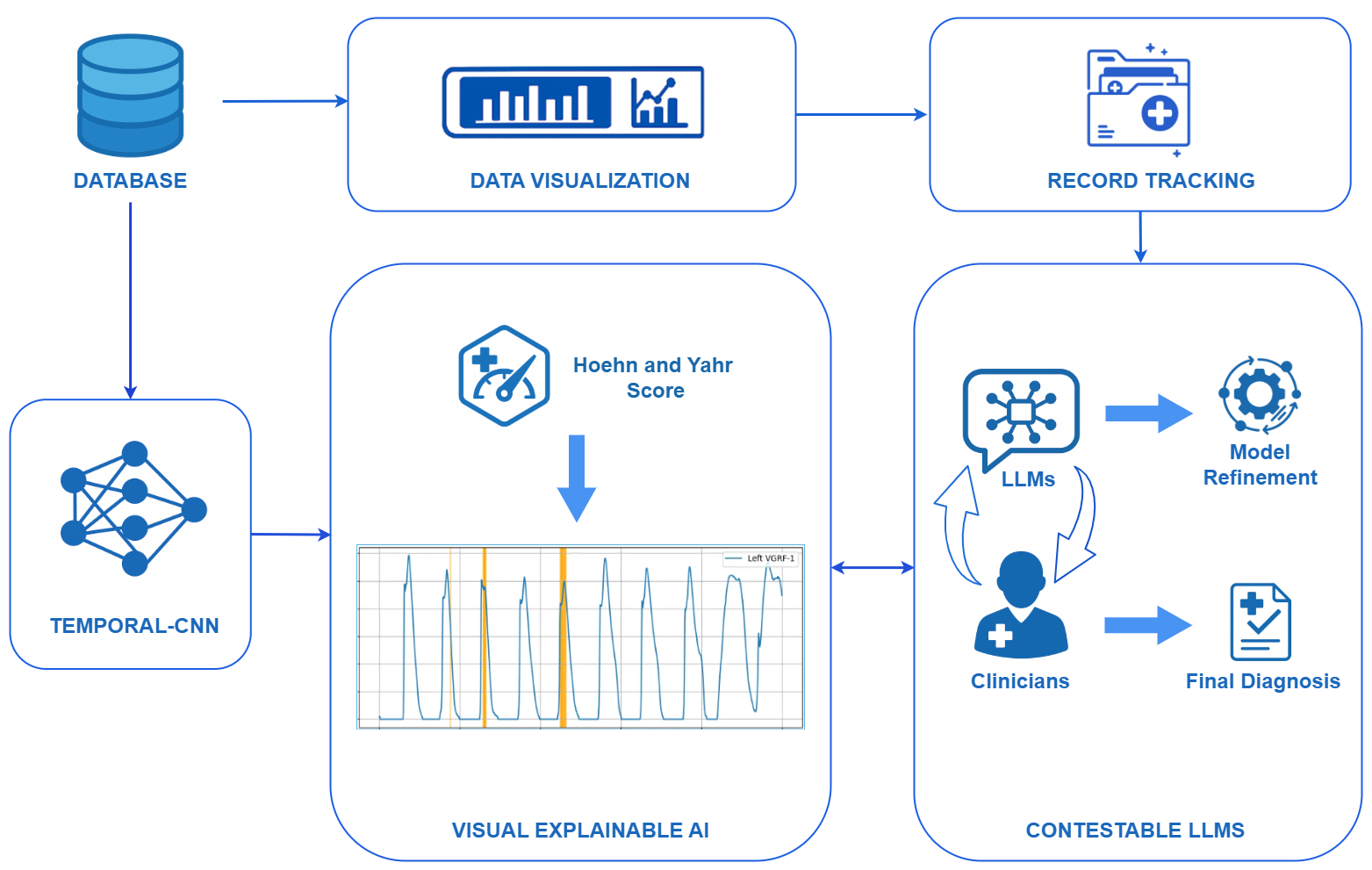}
    \caption{Overview of the proposed framework}
    \label{fig:overall}
\end{figure}

We introduce ConGaIT (Contestable Gait Interpretation \& Tracking), a clinician-centered dashboard that embeds CAI principles across all interaction layers. Rooted in HCI and procedural justice, the framework supports explainable, justifiable, and auditable AI use in PD care through a modular interface aligned with clinical and regulatory standards. An overview of the proposed framework is presented in Figure~\ref{fig:overall}.

\begin{figure}[t!]
    \centering
    \includegraphics[width=1\linewidth]{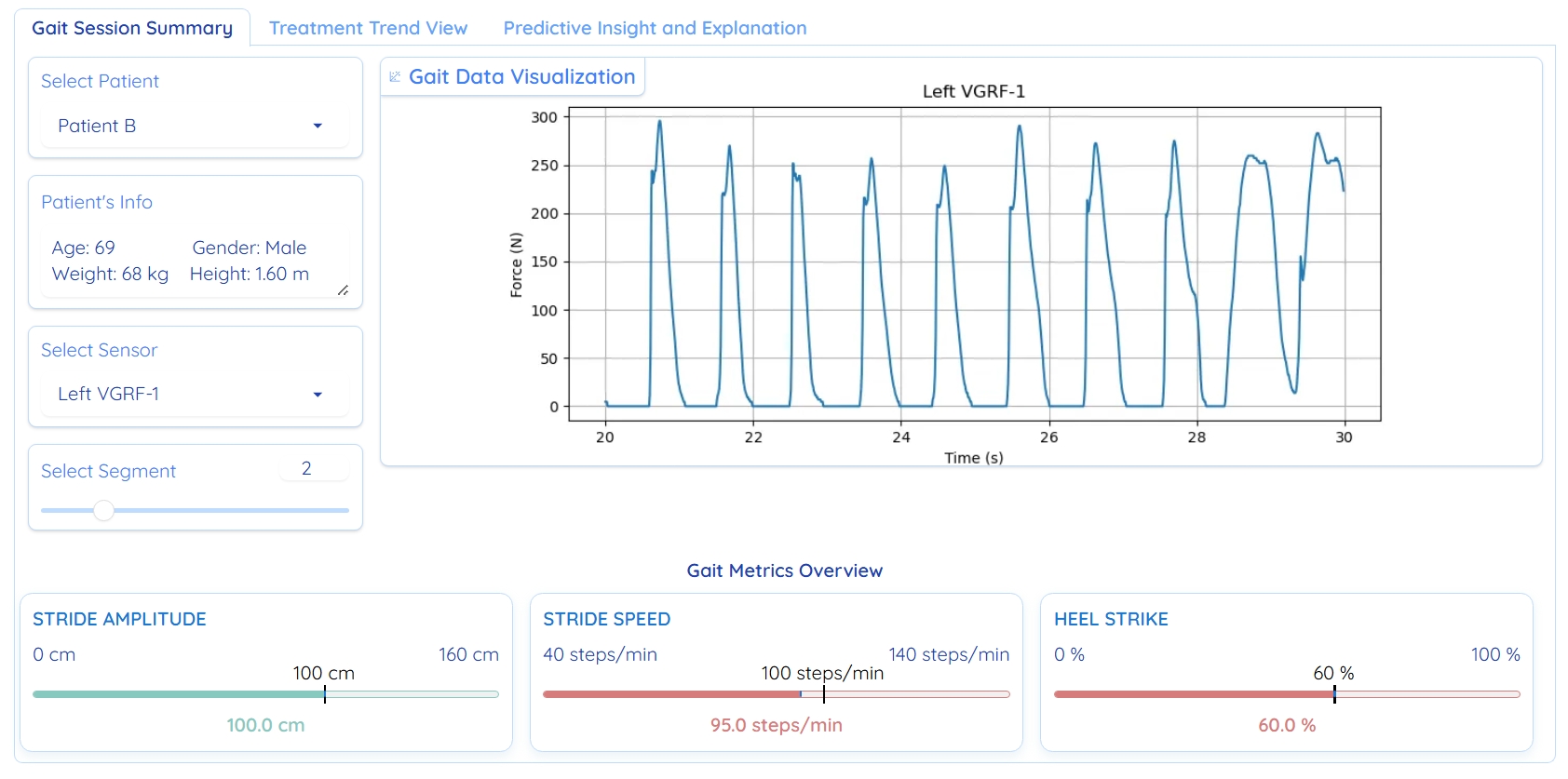}
    \caption{Overview of the Gait Session Summary tab}
    \label{fig:example1}
\end{figure}

\begin{figure}[t!]
    \centering
    \includegraphics[width=1\linewidth]{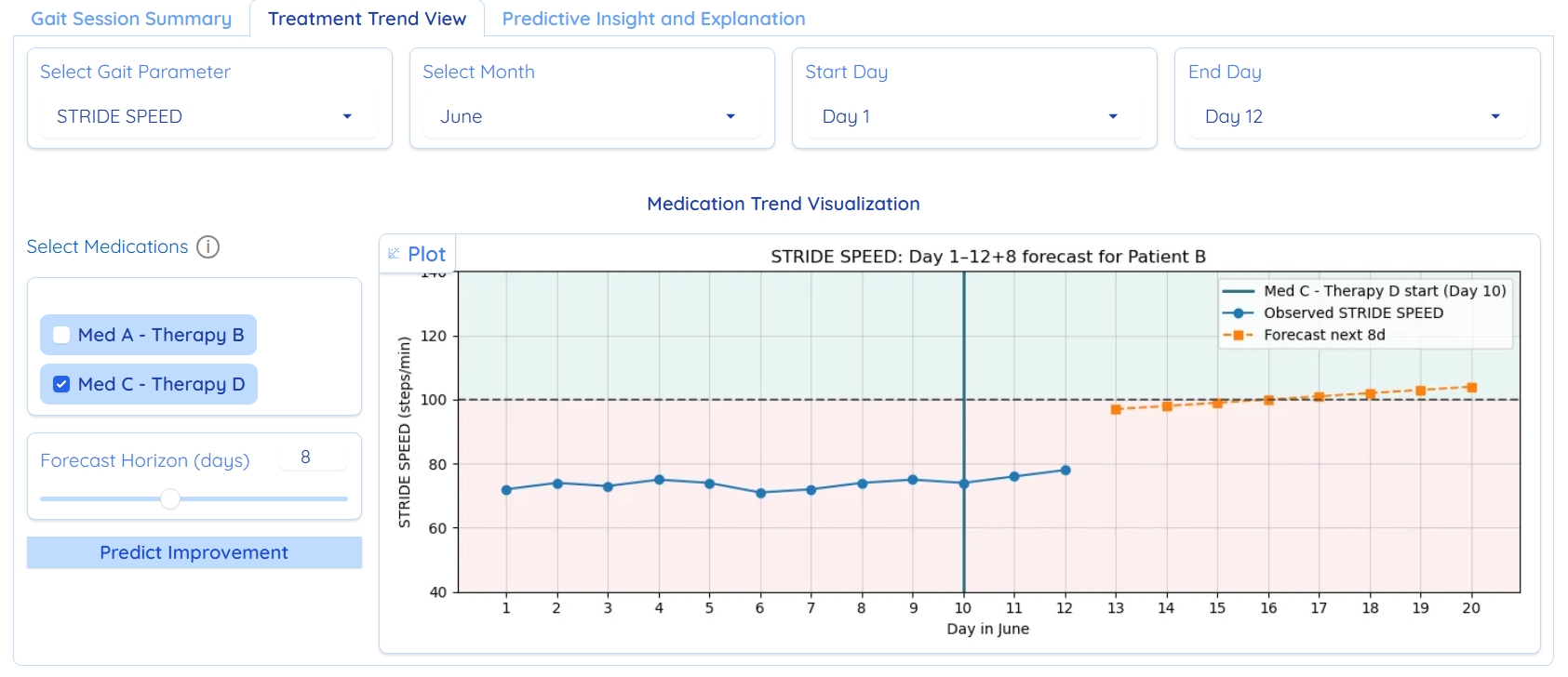} 
    \caption{Overview of the Treatment Trend View tab}
    \label{fig:example2}
\end{figure}

\begin{figure}[t!]
    \centering
    \includegraphics[width=1\linewidth]{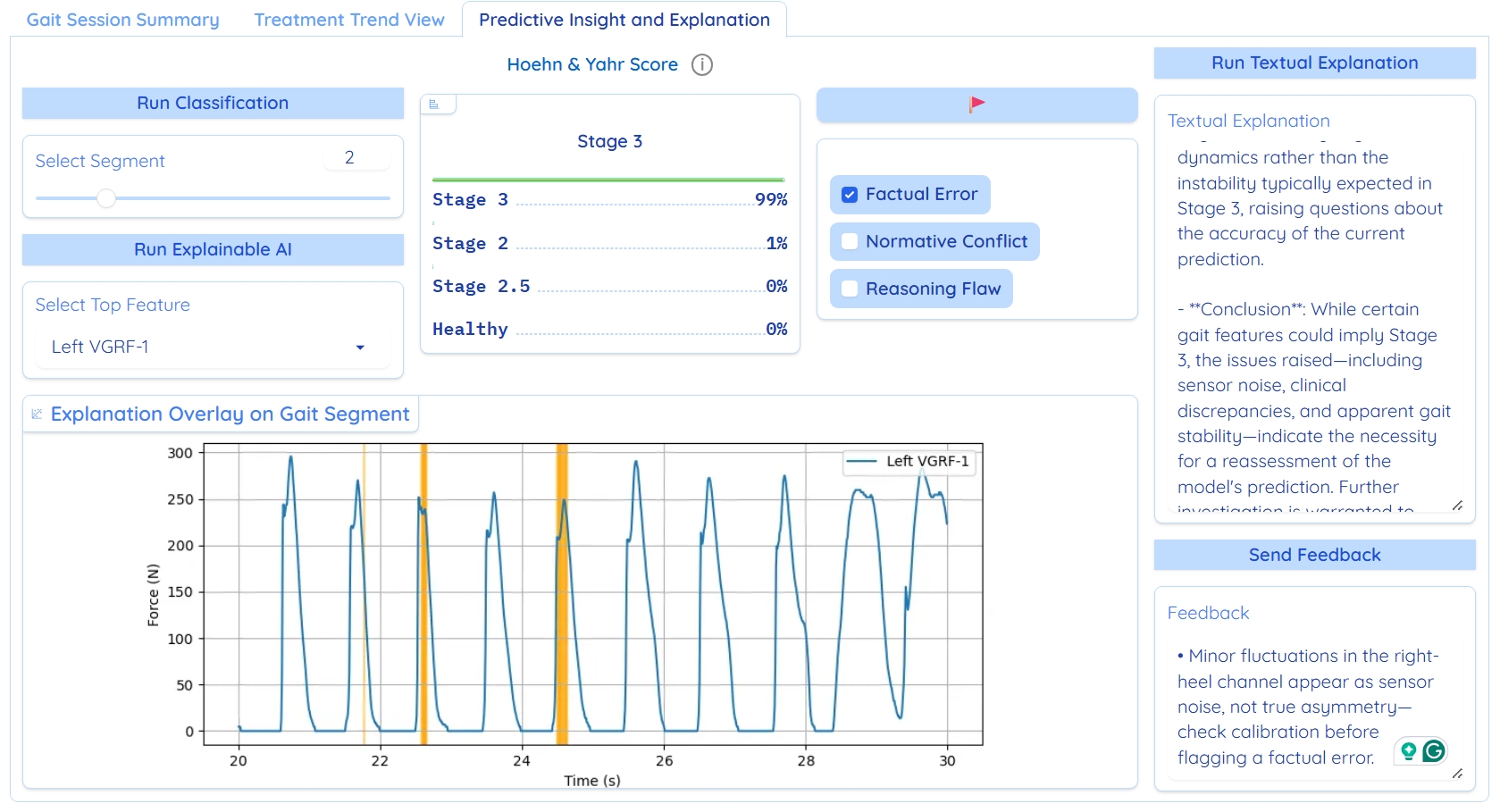}
    \caption{Overview of the Predictive Insight and Explanation tab}
    \label{fig:example3}
\end{figure}

ConGaIT comprises three core tabs (Figures~\ref{fig:example1}-\ref{fig:example3}). The \textbf{Gait Session Summary} visualizes per-session gait features (e.g., stride amplitude, freezing) using color-coded indicators aligned with normative data. Clinicians can interactively explore 10-second intervals, toggle sensor channels, and view raw vertical ground reaction force (VGRF) waveforms with context-aware tooltips. The \textbf{Treatment Trend View} tracks longitudinal gait changes alongside medication history, enabling overlay of interventions and use of AI-based forecasts to support treatment planning. Both tabs prioritize interpretability, traceability, and alignment with therapeutic decisions.

The \textbf{Predictive Insight and Explanation} tab implements core CAI functions. A convolutional neural network (CNN) classifier predicts Hoehn and Yahr stages \cite{hoehn1967parkinsonism} from 10-second gait windows, with Layer-wise Relevance Propagation (LRP) \cite{lapuschkin2016toolbox,slijepcevic2021explaining} identifying key sensors and time segments. A vision language model (OpenAI GPT-4o \cite{hurst2024gpt}) then generates justifications based on clinical rules. When a prediction is questionable, clinicians can initiate a \textit{Contest and Justify} flow by selecting an argument type: a Factual Error indicates incorrect input, a Normative Conflict reflects a mismatch with clinical context, and a Reasoning Flaw signals implausible attribution. The system responds with a justification that can be accepted or contested. All interactions are logged in an immutable audit trail to support oversight and model refinement.

\section{Evaluation and Discussion}
We applied the CAS \cite{moreira2025explainable} to evaluate the ConGaIT framework. The predictive backend was trained on the PhysioNet Gait in Parkinson’s Disease dataset \cite{goldberger2000physiobank} (93 patients, 73 controls) and achieves 96.58\% test accuracy. Observable criteria, such as Traceability and Auditing, were scored through direct inspection, while subjective criteria like Explanation Quality were assessed using simulated feedback from three clinician personas generated via OpenAI GPT-4 \cite{achiam2023gpt} and Microsoft TinyTroupe \cite{tinytroupe}: a general neurologist, a senior specialist, and a research clinician. These structured personas enabled reproducible, expert-informed evaluation aligned with CAS guidelines.

ConGaIT’s architecture is closely aligned with contestability goals, combining structured disagreement through the Contest and Justify interface with language model–generated justifications that enhance reasoning transparency. Features such as dual explanations, immutable logs, and structured argument types support adaptation, oversight, and institutional review. Scoring (shown in Table~\ref{tab:cas-congait}) reflects this integration: ConGaIT earned full marks for Explainability and Openness to Contestation, high scores in Traceability and Adaptivity, and full credit for Safeguards and Auditing. Clinician personas rated the interface 9 out of 10 for Ease of Contestation and 42 out of 50 for Explanation Quality, yielding a final CAS of 0.970. These results demonstrate that ConGaIT provides a practical framework for contestable clinical AI, introduces a novel HCI pattern for precise and auditable disagreement, and benefits from objective validation through CAS. While synthetic personas ensure reproducibility, future work will focus on clinical usability testing and fine-tuning the justification engine to enhance reasoning depth and regulatory alignment.

\begin{table}[ht!]
\centering
\caption{Evaluation results of ConGaIT}
\label{tab:cas-congait}
\begin{adjustbox}{width=0.46\textwidth}
\begin{tabular}{@{}lcccc@{}}
\toprule
\textbf{Property} & \textbf{Max} & \textbf{Weight ($\lambda$)} & \textbf{Score ($s_p$)} & \textbf{Final CAS} \\
\midrule
Explainability            & 2   & 0.30 & 2   & 0.300 \\
Openness to Contestation  & 2   & 0.12 & 2   & 0.120 \\
Traceability              & 10  & 0.12 & 9   & 0.108 \\
Built-in Safeguards       & 1   & 0.12 & 1   & 0.120 \\
Adaptivity                & 2   & 0.10 & 2   & 0.100 \\
Auditing                  & 2   & 0.10 & 2   & 0.100 \\
Ease of Contestation      & 10  & 0.07 & 9   & 0.063 \\
Explanation Quality       & 50  & 0.07 & 42  & 0.059 \\
\midrule
\textbf{Total CAS}        &     & \textbf{1.00} &       & \textbf{0.970} \\
\bottomrule
\end{tabular}
\end{adjustbox}
\end{table}

\section{Conclusion}
This study presented ConGaIT, a clinician-centered dashboard that integrates CAI principles into PD gait monitoring. By combining visual and textual explanations, structured contestation workflows, and immutable audit trails, the system enhances transparency, user agency, and regulatory alignment. Evaluation using the CAS yielded a high score of 0.970, confirming its effectiveness across human-centered, technical, legal, and organizational dimensions. Through a novel proxy evaluation with language model–generated clinician personas, we demonstrated that contestability can be embedded early in the design process. ConGaIT offers a practical and reproducible blueprint for implementing responsible, contestable AI in high-stakes clinical settings.

\bibliographystyle{ieeetr}
\bibliography{ref}

\end{document}